\documentclass[aps,pre,a4paper,twocolumn,showkeys,floatfix]{revtex4-1}
\usepackage[T1]{fontenc}
\usepackage{amsmath}
\usepackage{amssymb}
\usepackage{graphicx}
\usepackage{psfrag}
\usepackage{csquotes}
\usepackage{xcolor}
\usepackage{listings}
\usepackage{bibspacing}
\usepackage[colorlinks=true,hyperfootnotes=true,breaklinks=true,backref=none]{hyperref}
\setcitestyle{authoryear,round,semicolon,aysep={,}}

\begin{document}
\title{Paradox of integration---Dynamics of two-dimensional status}

\author{K.~Malarz}
\homepage{http://home.agh.edu.pl/malarz/}
\email{malarz@agh.edu.pl}
\affiliation{\href{http://www.agh.edu.pl/}{AGH University of Science and Technology},
\href{http://www.pacs.agh.edu.pl/}{Faculty of Physics and Applied Computer Science},\\
al. Mickiewicza 30, 30-059 Krakow, Poland.}

\author{K.~Ku{\l}akowski}
\email{Krzysztof.Kulakowski@fis.agh.edu.pl}
\affiliation{\href{http://www.agh.edu.pl/}{AGH University of Science and Technology},
\href{http://www.pacs.agh.edu.pl/}{Faculty of Physics and Applied Computer Science},\\
al. Mickiewicza 30, 30-059 Krakow, Poland.}


\begin{abstract}
According to Peter M.~\citet[Exchange and Power in Social Life, Wiley and Sons,][p. 43]{blau1}, the process of integration of a newly formed group
has a paradoxical aspect: most attractive individuals are rejected because they raise fear of rejection. Often, their solution is 
to apply a self-deprecating strategy, which artificially raises the social statuses of their opponents. Here we introduce a two-dimensional 
space of status, and we demonstrate that with this setup, the self-deprecating strategy efficiently can prevent the rejection. Examples of application of this 
strategy in the scale of a society are provided.
\end{abstract}

\date{\today}

\maketitle

%% ############################################################################
\section{Introduction}
%% ############################################################################

To say that social processes are complex would be an understatement. Such is also the dynamics of social integration of a newly formed
group, where social statuses of individuals are formed during multiple encounters in the presence of culturally grounded beliefs \citep{clr}.
According to Peter M. \citet{blau1}, the process of status formation can be divided in two stages. The first stage is just attempts to gain high status
by demonstrating actors' chief assets. Once this is accomplished, in the second stage the actors with high status apply a self-deprecating
strategy (SDS) to evade fear-driven rejection by those with lower status. As \citet[p. 43]{blau1} puts it: `{\em In a group situation, impressive qualities
make a person attractive in one sense and unattractive in another,
because they raise fears of rejection and pose a status threat for the rest of the group}'. Blau termed this
combination of attraction and repulsion as `paradox of integration'. SDS consists in revealing minor defects of her/himself and self-mockery, but
also in displaying of friendly interest in contact with less attractive neighbors. It should be noted that the condition of an application of SDS is
that the status difference should be maintained. In an experimental situation,
persons with high status maintained differences in opinions important for their status, but `{\em had a somewhat greater tendency to conform
to the opinions of subordinates on miscellaneous topics}' \cite[p. 54]{blau1}. This tendency is an example of SDS. On the contrary,
SDS used by non-attractive persons turns out to be boring for the rest of the group \citep{blau1,goff}.

According to Max \citet{weber}, social status is one of main three `different but interrelated' \citep{clr} origins of social inequalities,
along with power and resources. The status itself is known to be multidimensional, as it comprises ethnicity, gender, age, occupational group,
education, class-based lifestyle etc. On the other hand, as noted by Cecilia \citet{clr2000} in the context of gender inequalities `{\em ...on the things
that really count in our society, competence and agency, the advantaged group is seen as better. But on positive attributes of secondary value
in the dominant culture, the disadvantaged group is allowed to be better}'.

For purposes of our analysis, the split between status
characteristics should be made between relevant and less relevant ones, and the detailed list of the content of both sets is not that crucial.
This assertion finds a confirmation in an experimental research, where an influence of nominal attributes has been found to produce the same effect
as a pay level \citep{clr2}.

In \citet{m1,m2} previous approach, the model has been formulated as follows. The network structure was a complete graph;
each actor could interact with each other. The interaction was directed; actor X was praising or criticizing actor Y.
The probability of this or that decision depended on the status of X and on the number of actors with the same status as the one of Y:
those more numerous were criticized less willingly, and those with higher status were praising more frequently. Both these assumptions reflect
the aim of SDS: to evade rejection by numerous group fraction by praising those of lower status. The drawback of
this model was that the SDS was not efficient as a defense; the sympathy of agent X to Y was a dependent
variable which did not influence the above probabilities. Here we present an attempt to remove this drawback by introducing
a two-dimensional space of status, say $A$ and $B$ (real and surface status, respectively). Here SDS is to increase the status $B$ of persons
with lower status $A$. Simultaneously we assume that the fear driven rejection, which lowers the status $A$ of the target actor, is active only
against actors with higher status $A$ and the same status $B$.

Similarly to \citet{m2} work, the distribution of actors in the space of status is used here, and not individual actors. This choice is motivated by
indications, that the process of status distribution is active both in micro- and macro-level, abundant in the literature \citep{clr,clr3}. However,
here we do not apply the mean-field decoupling of praising or not and being praised or not, which was used by \citet{m2}.

%% ############################################################################
\section{Dynamics of two-dimensional status}
%% ############################################################################

There are two processes: those who have the same $B$-status and lower $A$-status intend to reduce the difference (fear-induced rejection), and those
who have the same $B$-status and higher $A$-status intend to reduce the conflict by enhancement of $B$-status of their opponents (SDS). All other processes are set to zero. In other words, there is no interaction between actors in the same state $(A,B)$ and there is
no interaction between actors with different status $B$. The latter can be justified as follows: on the contrary to the status $A$, the status $B$
can be varied practically without limits and at little cost by building large multilevel hierarchies, as medals of numerous kinds.

The above process may be described by the master equation 
\begin{widetext}
\begin{eqnarray}
	\dfrac{dv(A,B)}{dt}=&-&v(A,B)\big[w(A\to A-1,B)+w(A\to A+1,B)+w(A,B\to B-1)+w(A,B\to B+1)\big]\nonumber\\
	&+&v(A+1,B)w(A+1 \to A,B)+v(A-1,B)w(A-1 \to A,B)\nonumber\\
	&+&v(A,B+1)w(A,B+1\to B)+v(A,B-1)w(A,B-1\to B),
\label{me2}
\end{eqnarray}
\end{widetext}
where $v(A,B)$ is the density of actors of status ($A,B$), and $w(A\to A',B)$, $w(A,B\to B')$ are the rates of the related processes.
The normalization condition is
\begin{equation}
	\sum_{A,B=-\infty}^{+\infty}v(A,B)=1.
	\label{norm2}
	\end{equation}

For the rates, the formulas as follows can be proposed:
\begin{subequations}
\label{rates2}
\begin{eqnarray}
	w(A\to A-1,B) &=&\alpha\sum_{A'=-\infty}^{A-1} v(A',B);\\
	w(A\to A+1,B) &=&0;\nonumber\\
	w(A,B\to B-1) &=&0;\nonumber\\
	w(A,B\to B+1) &=&\beta\sum_{A'=A+1}^{\infty} v(A',B);\\
	w(A+1\to A,B) &=&\alpha\sum_{A'=-\infty}^{A} v(A',B);\\
	w(A-1\to A,B) &=&0;\nonumber\\
	w(A,B+1\to B) &=&0;\nonumber\\
	w(A,B-1\to B) &=&\beta\sum_{A'=A+1}^{\infty} v(A',B-1).
\end{eqnarray}
\end{subequations}
As we see, the conflict is going on along the axis of the $A$-status. The rates proposed mean that all the actors with the $A$-status $A'$ lower than $A$ contribute to rejection, and all the actors with the $A$-status $A'$ higher than $A$ contribute to SDS. The coefficients $\alpha$ and $\beta$ measure the intensity of rejection and SDS, respectively.

%% ############################################################################
\section{Calculations}
%% ############################################################################

%% ----------------------------------------------------------------------------
\begin{video}[t]
\psfrag{A}{$A$}
\psfrag{B}{$B$}
\psfrag{v(A,B)}{$v(A,B)$}
\psfrag{t=0}{$t=0$}
\psfrag{t=5}{$t=5$}
\psfrag{t=10}{$t=10$}
\psfrag{t=20}{$t=20$}
\psfrag{t=50}{$t=50$}
\psfrag{t=100}{$t=100$}
\psfrag{t=200}{$t=200$}
\psfrag{t=500}{$t=500$}
\psfrag{t=1000}{$t=1000$}
\includegraphics[width=0.95\columnwidth]{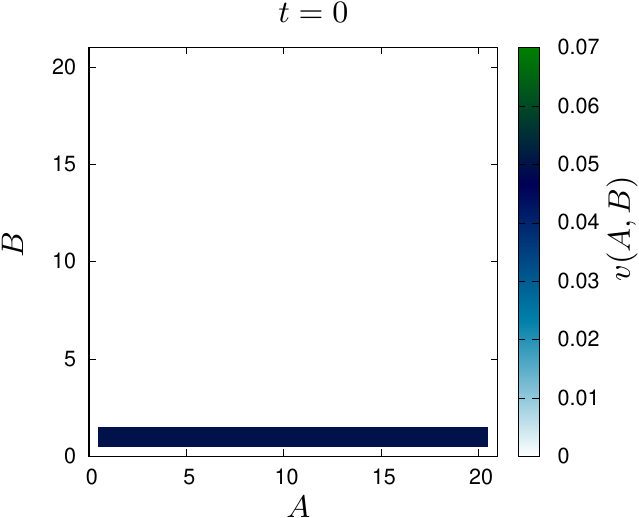}\\
\includegraphics[width=0.95\columnwidth]{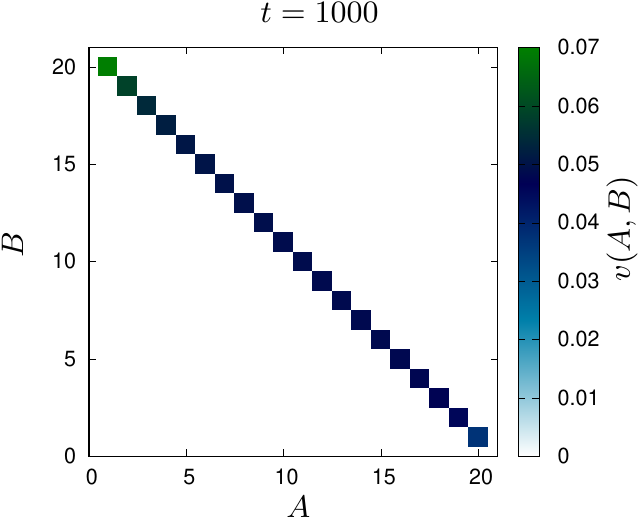}
\setfloatlink{http://www.zis.agh.edu.pl/files/hierar_Lx20_Ly20_alpha005.gif}
\caption{\label{vid:hierar_Lx20_Ly20_alpha005} The spatial--temporal evolution of the density of actors status $v(A,B)$ for the hierarchical initial state and $\alpha=0.05$, $\beta=1-\alpha$, $L_A=L_B=20$, $\delta t=10^{-5}$}
\end{video}
%% ----------------------------------------------------------------------------
%% ----------------------------------------------------------------------------
\begin{video}[t]
\psfrag{A}{$A$}
\psfrag{B}{$B$}
\psfrag{v(A,B)}{$v(A,B)$}
\psfrag{t=0}{$t=0$}
\psfrag{t=5}{$t=5$}
\psfrag{t=10}{$t=10$}
\psfrag{t=20}{$t=20$}
\psfrag{t=50}{$t=50$}
\psfrag{t=100}{$t=100$}
\psfrag{t=200}{$t=200$}
\psfrag{t=500}{$t=500$}
\psfrag{t=1000}{$t=1000$}
\includegraphics[width=0.95\columnwidth]{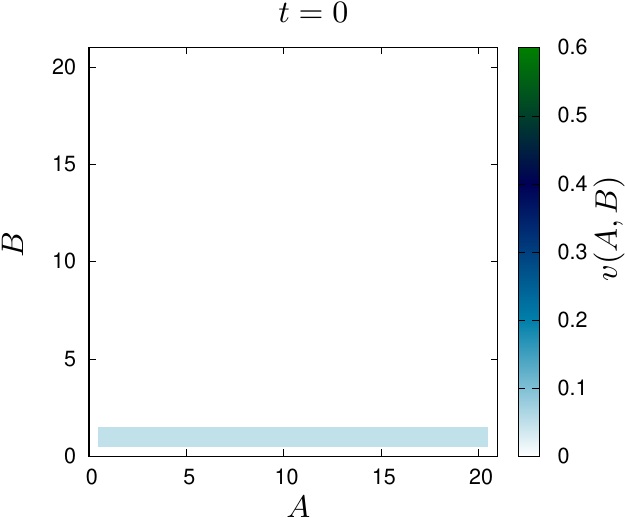}\\
	\includegraphics[width=0.95\columnwidth]{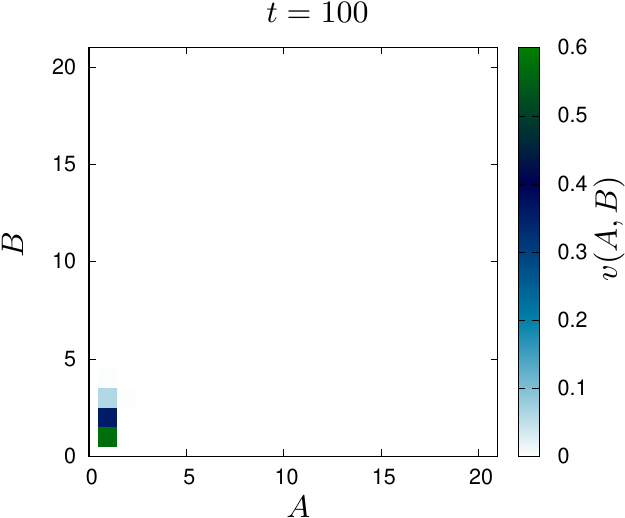}
\setfloatlink{http://www.zis.agh.edu.pl/files/hierar_Lx20_Ly20_alpha095.gif}
\caption{\label{vid:hierar_Lx20_Ly20_alpha095}The spatial--temporal evolution of the density of actors status $v(A,B)$ for the hierarchical initial state and $\alpha=0.95$, $\beta=1-\alpha$, $L_A=L_B=20$, $\delta t=10^{-5}$}
\end{video}
%% ----------------------------------------------------------------------------
%% ----------------------------------------------------------------------------
\begin{video}[b]
\psfrag{A}{$A$}
\psfrag{B}{$B$}
\psfrag{v(A,B)}{$v(A,B)$}
\psfrag{t=0}{$t=0$}
\psfrag{t=5}{$t=5$}
\psfrag{t=10}{$t=10$}
\psfrag{t=20}{$t=20$}
\psfrag{t=50}{$t=50$}
\psfrag{t=100}{$t=100$}
\psfrag{t=200}{$t=200$}
\psfrag{t=500}{$t=500$}
\psfrag{t=1000}{$t=1000$}
\includegraphics[width=0.95\columnwidth]{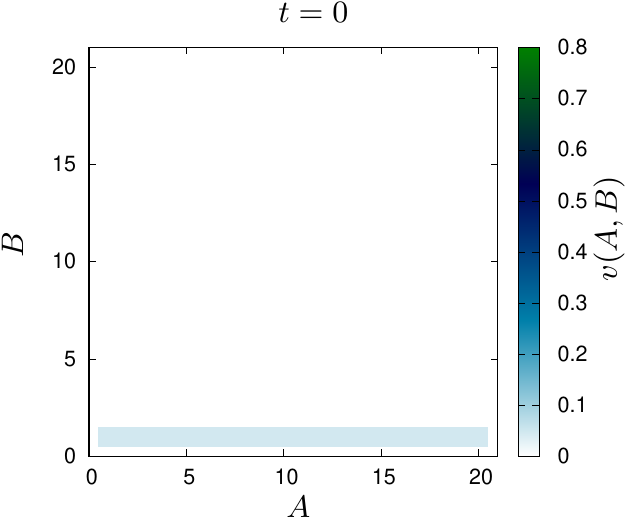}\\
	\includegraphics[width=0.95\columnwidth]{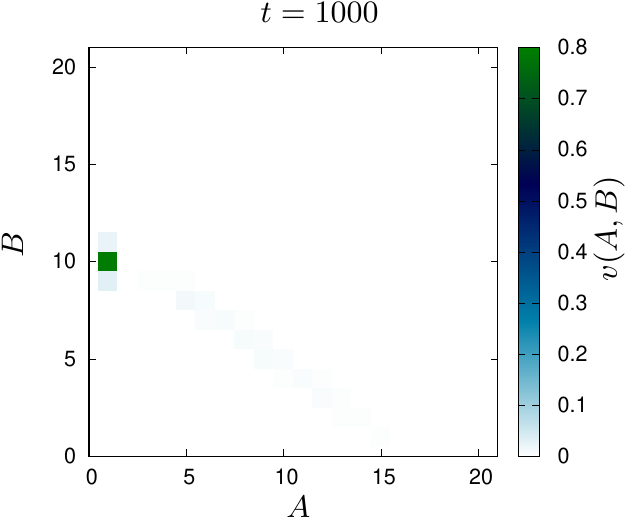}
\setfloatlink{http://www.zis.agh.edu.pl/files/hierar_Lx20_Ly20_alpha050.gif}
\caption{\label{vid:hierar_Lx20_Ly20_alpha050} The spatial--temporal evolution of the density of actors status $v(A,B)$ for the hierarchical initial state and $\alpha=0.5$, $\beta=1-\alpha$, $L_A=L_B=20$, $\delta t=10^{-5}$}
\end{video}
%% ----------------------------------------------------------------------------

%% ----------------------------------------------------------------------------
%% ini=homogenous, Lx=40, Ly=60
%% ----------------------------------------------------------------------------
\begin{video*}[t]
\psfrag{A}{$A$}
\psfrag{B}{$B$}
\psfrag{v(A,B)}{$v(A,B)$}
\psfrag{t=0}{$t=0$}
\psfrag{t=500}{$t=500$}
\psfrag{t=1000}{$t=1000$}
\psfrag{t=3000}{$t=3000$}
\includegraphics[height=0.23\textheight]{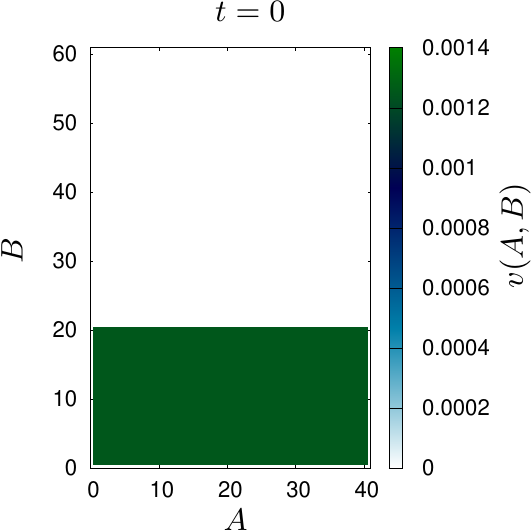}\hfill\includegraphics[height=0.23\textheight]{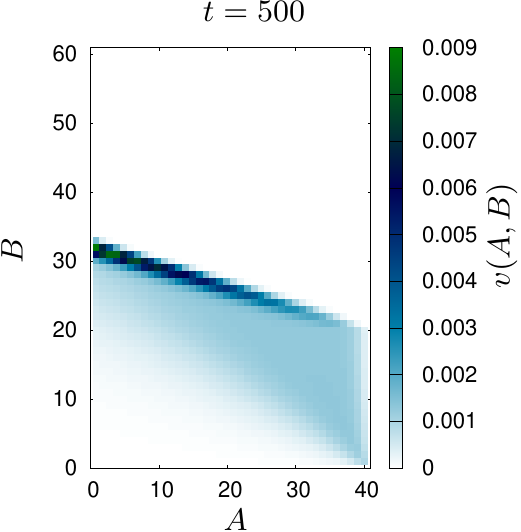}\hfill\includegraphics[height=0.23\textheight]{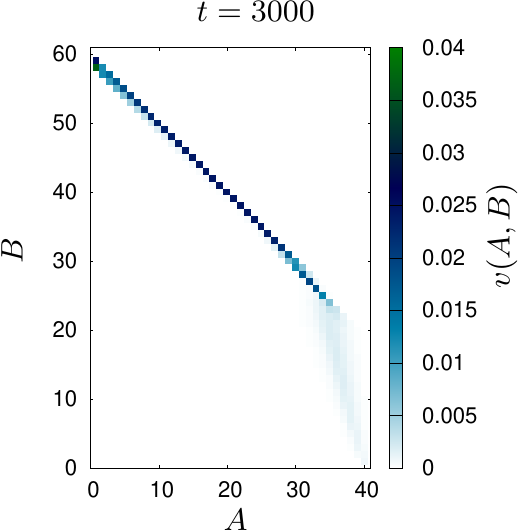}
\setfloatlink{http://www.zis.agh.edu.pl/files/homoge_Lx40_Ly60_alpha005.gif}
\caption{\label{vid:homogeLx40Ly60alpha005}The spatial--temporal evolution of the density of actors status $v(A,B)$ for homogenoues initial state and $\alpha=0.05$, $\beta=1-\alpha$, $L_A=40$, $L_B=60$, $\delta t=10^{-5}$}
\end{video*}
%% ----------------------------------------------------------------------------
\begin{video*}[t]
\psfrag{A}{$A$}
\psfrag{B}{$B$}
\psfrag{v(A,B)}{$v(A,B)$}
\psfrag{t=0}{$t=0$}
\psfrag{t=500}{$t=500$}
\psfrag{t=1000}{$t=1000$}
\psfrag{t=3000}{$t=3000$}
\includegraphics[height=0.23\textheight]{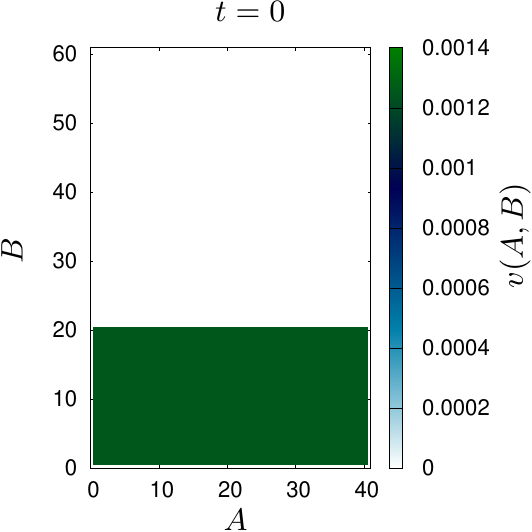}\hfill\includegraphics[height=0.23\textheight]{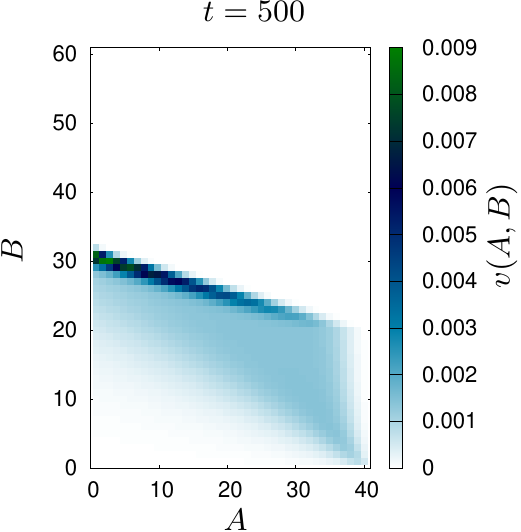}\hfill\includegraphics[height=0.23\textheight]{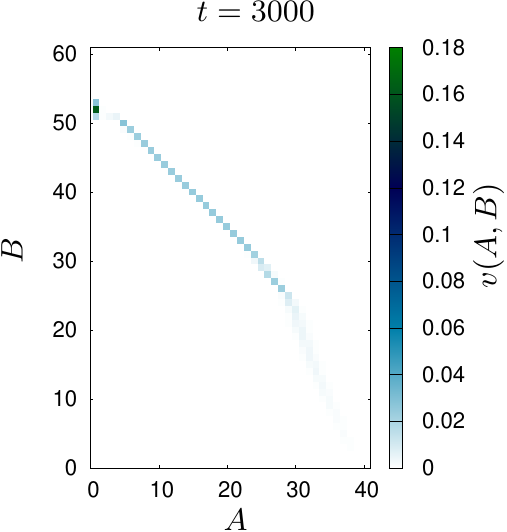}
\setfloatlink{http://www.zis.agh.edu.pl/files/homoge_Lx40_Ly60_alpha015.gif}
\caption{\label{vid:homogeLx40Ly60alpha015}The spatial--temporal evolution of the density of actors status $v(A,B)$ for homogenoues initial state and $\alpha=0.15$, $\beta=1-\alpha$, $L_A=40$, $L_B=60$, $\delta t=10^{-5}$}
\end{video*}
%% ----------------------------------------------------------------------------

It is clear from Eq.~\eqref{rates2} that the obtained multidimensional trajectory of $v(A,B)$ depends only on the ratio $\alpha/\beta$, then we
keep $\beta=1-\alpha$. On the other hand, the initial state does matter. In particular, once $A=A'$ for each pair of occupied cells, the whole state
is absorbing and does not evolve at all. We are particularly interested in an initial state where the hierarchy is set along $A$-axis. This is the
state when the first stage of the process is completed, according to the description of the status formation by Blau, as mentioned in the Introduction.
Also, some disordered initial configurations are explored here, as they are expected to be generic. These states are realized as radom distributions
of $v(A,B)$, localized on some subset of cells near the lattice centre. The logic of the theory dictates that the simulation
is performed on a rectangle piece of a lattice with open boundary conditions.
Basically, the lattice size $L_A\times (L_B+L_A)$ is sufficient, where $(L_A\times L_B)$ is only slightly larger than the size of the initial configuration.
This is because along $A$ axis, only contraction of the occupied area is expected. Mathematically,
the dynamic system \eqref{me2} is akin to a set of deterministic nonlinear integro--differential equations, with the only difference that the space
is discrete here. Both statuses $A$ and $B$ are expressed in ordinal variables, therefore the simulation clarifies the picture only within a
semi-quantitative model.

We apply Euler's method to solve Eq.~\eqref{me2} numerically \citep[p. 434]{Stoer1980}. 
The Euler's methods is the simplest method for solving initial-value problem
\begin{equation}
\label{eq:initial-value-problem}
\frac{dy}{dt}=f(t,y), \qquad y(t_0)=y_0,
\end{equation}
basing on approximation
\begin{equation}
\dfrac{y(t+\delta t)-y(t)}{\delta t} \approx f(t,y(t)),
\end{equation}
and thus
\begin{equation}
y(t+\delta t)\approx y(t)+\delta t \cdot f(t,y(t)),
\end{equation}
which yields iterative scheme for finding approximate solution $\eta_i=\eta(t_i)$ of Eq.~\eqref{eq:initial-value-problem} at equidistant points $t_i=t_0+i\delta t$ as
\begin{subequations}
\begin{eqnarray}
	& &\eta_0\equiv y_0;\\
	& &\eta_{i+1}\equiv \eta_i + \delta t\cdot f(t_i,\eta_i)\text{ for } i=0, 1, 2,\ldots
\end{eqnarray}
\end{subequations}
Differential scheme for solving Eq.~\eqref{me2} is then:
\begin{widetext}
\begin{eqnarray}
\label{eq:Euler}
	v(A,B;t+\delta t)&=&v(A,B;t)\nonumber\\
			 &+&\delta t\big( -v(A,B;t)[w(A\to A-1,B)+w(A,B\to B+1)]\nonumber\\
			 &+&v(A+1,B;t)w(A+1\to A,B)+v(A,B-1;t)w(A,B-1\to B) \big)
\end{eqnarray}
\end{widetext}
applied synchronously to all sites of the grid $\mathcal{G}=\{(A,B):1\le A\le L_A, 1\le B\le L_B, A\in\mathbb{Z}, B\in\mathbb{Z}\}$.
The impementation of the method \eqref{eq:Euler} and various initial values $v(A,B;t=0)$ (defined in Secs.~\ref{sec:Hierarchical}--\ref{sec:Random}) are displayed in Listing~\ref{lst:main}.
We have checked numerically, that the scheme \eqref{eq:Euler} is good enough for keeping sum \eqref{norm2} constant during simulation for $\delta t=10^{-5}$ and $t\le 3000$, i.e. for $3\cdot10^8$ time steps.

%% ############################################################################
\section{Results}
%% ############################################################################

%% ----------------------------------------------------------------------------
\begin{video*}[t]
\psfrag{A}{$A$}
\psfrag{B}{$B$}
\psfrag{v(A,B)}{$v(A,B)$}
\psfrag{t=0}{$t=0$}
\psfrag{t=500}{$t=500$}
\psfrag{t=1000}{$t=1000$}
\psfrag{t=3000}{$t=3000$}
\includegraphics[height=0.23\textheight]{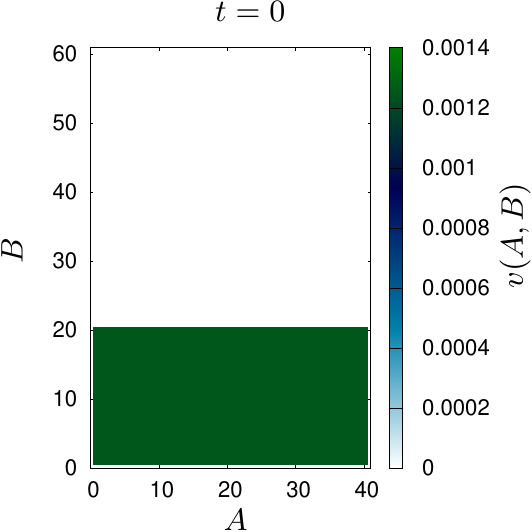}\hfill\includegraphics[height=0.23\textheight]{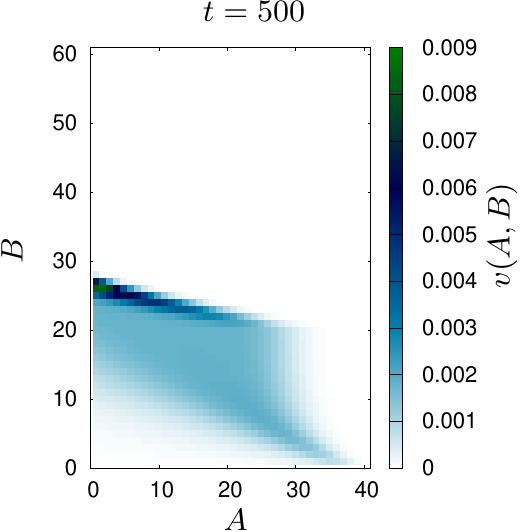}\hfill\includegraphics[height=0.23\textheight]{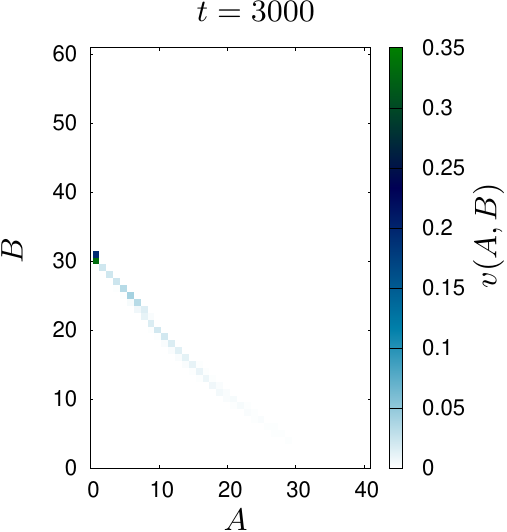}
\setfloatlink{http://www.zis.agh.edu.pl/files/homoge_Lx40_Ly60_alpha050.gif}
\caption{\label{vid:homogeLx40Ly60alpha050}The spatial--temporal evolution of the density of actors status $v(A,B)$ for homogenoues initial state and $\alpha=0.5$, $\beta=1-\alpha$, $L_A=40$, $L_B=60$, $\delta t=10^{-5}$}
\end{video*}
%% ----------------------------------------------------------------------------
\begin{video*}[t]
\psfrag{A}{$A$}
\psfrag{B}{$B$}
\psfrag{v(A,B)}{$v(A,B)$}
\psfrag{t=0}{$t=0$}
\psfrag{t=500}{$t=500$}
\psfrag{t=1000}{$t=1000$}
\psfrag{t=3000}{$t=3000$}
\includegraphics[height=0.23\textheight]{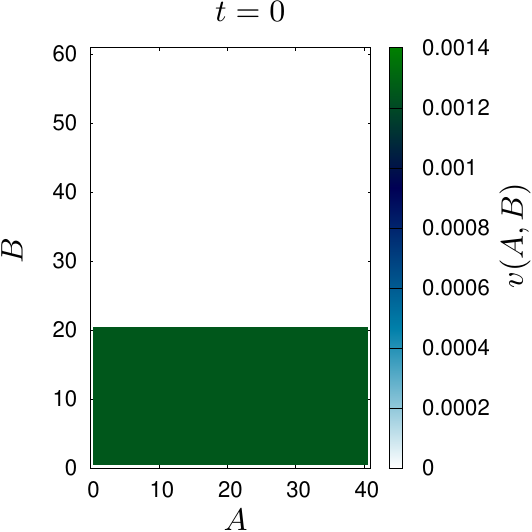}\hfill\includegraphics[height=0.23\textheight]{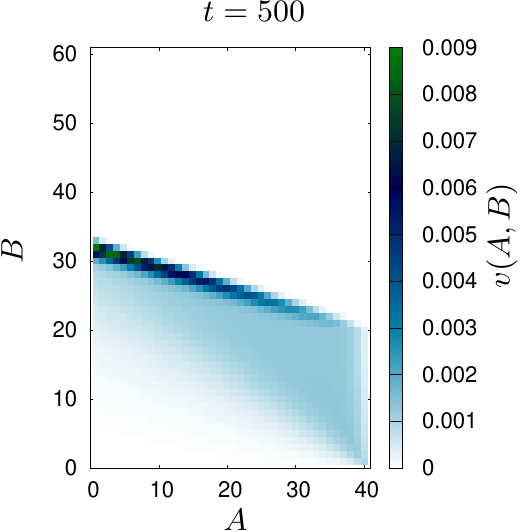}\hfill\includegraphics[height=0.23\textheight]{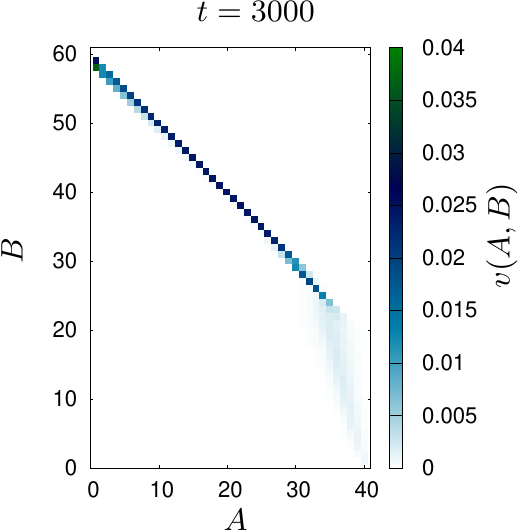}
\setfloatlink{http://www.zis.agh.edu.pl/files/homoge_Lx40_Ly60_alpha095.gif}
\caption{\label{vid:homogeLx40Ly60alpha095}The spatial--temporal evolution of the density of actors status $v(A,B)$ for homogenoues initial state and $\alpha=0.95$, $\beta=1-\alpha$, $L_A=40$, $L_B=60$, $\delta t=10^{-5}$}
\end{video*}
%% ----------------------------------------------------------------------------

%% ============================================================================
\subsection{\label{sec:Hierarchical}Hierarchical initial state}
%% ============================================================================

For the hierarchical initial state, the line of $L_A$ cells along $A$ axis ($B_0=\text{const}$) is occupied with weights $v(A,B_0)=1/L_A$. The final (absorbing) configurations depend on the ratio $\alpha/\beta$. The cases $\alpha=0.05$ and $\alpha=0.95$ are shown in Videos~\ref{vid:hierar_Lx20_Ly20_alpha005} and~\ref{vid:hierar_Lx20_Ly20_alpha095}, respectively. For $\alpha=0.05$ SDS prevails, and the rejection is very weak. What we observe is just a reorientation of the axis of
hierarchy. At the end of evolution, those with minimal status $A$ get the maximal status $B$. On the square lattice, the final slope of the axis
is necessarily equal to $3\pi /4$. However, we do not claim that we dispose a common scale of units of status $A$ vs $B$; this angle is just the result of the parameterization chosen here.

For $\alpha=0.95$, SDS is almost invisible, as the rejection reduces the scale $A$ to almost one cell---statuses of all are equal in both dimensions. The results obtained for an intermediate case $\alpha=\beta=0.5$ are shown in Video~\ref{vid:hierar_Lx20_Ly20_alpha050}. In any case, in the absorbing state there is no occupied cells with the same status $A$ and different status $B$. 

%% ============================================================================
\subsection{\label{sec:Homogenous}Homogenous initial state}
%% ============================================================================

For a homogenous initial state and small values of the rate $\alpha$, both effects are visible (see Videos~\ref{vid:homogeLx40Ly60alpha005} and \ref{vid:homogeLx40Ly60alpha015}): the reduction of the $A$ coordinate and the enhancement of the $B$ coordinate. These effects mutually influence each other; the former is the strongest for largest $A$, as it is driven by all cells with smaller $A$ and the same $B$. As the number of those actors decrease with the rate $\beta$, the reduction of $A$ for the largest $A$ is slowed down. On the other hand, the rate of the enhancement of $B$ decreases as well. Finally,the absorbing configuration is the same as for the hierarchical initial state; a diagonal straight line of cells, with the largest $B$ appears for the smallest $A$.

For larger rate $\alpha \ge 0.5$ the rejection dominates, and the whole system is reduced to one cell of minimal $A$ (see Videos~\ref{vid:homogeLx40Ly60alpha050} and \ref{vid:homogeLx40Ly60alpha095}).

%% ============================================================================
\subsection{\label{sec:Random}Random initial state}
%% ============================================================================

For a random initial state, the final results are approximately the same as for the homogeneous initial state. The difference is that during some 
transient time, some remains of the inhomogeneities are visible (see Videos~\ref{vid:randomLx40Ly60alpha005}--\ref{vid:randomLx40Ly60alpha095}).

%% ----------------------------------------------------------------------------
%% ini=random, Lx=40, Ly=60
%% ----------------------------------------------------------------------------
\begin{video*}
\psfrag{A}{$A$}
\psfrag{B}{$B$}
\psfrag{v(A,B)}{$v(A,B)$}
\psfrag{t=0}{$t=0$}
\psfrag{t=500}{$t=500$}
\psfrag{t=1000}{$t=1000$}
\psfrag{t=3000}{$t=3000$}
\includegraphics[height=0.23\textheight]{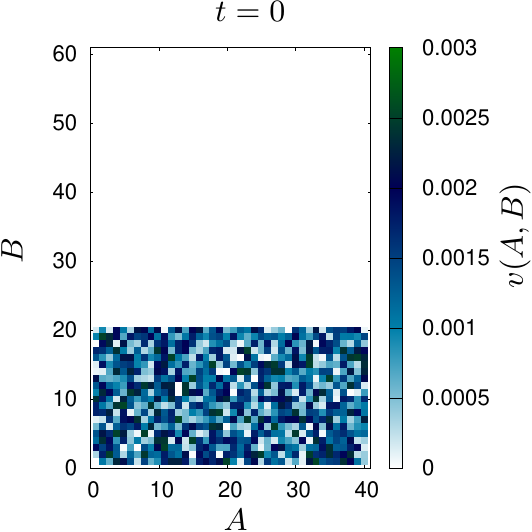}\hfill\includegraphics[height=0.23\textheight]{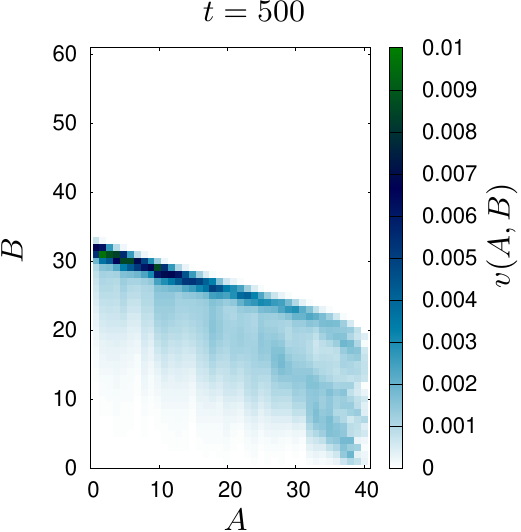}\hfill\includegraphics[height=0.23\textheight]{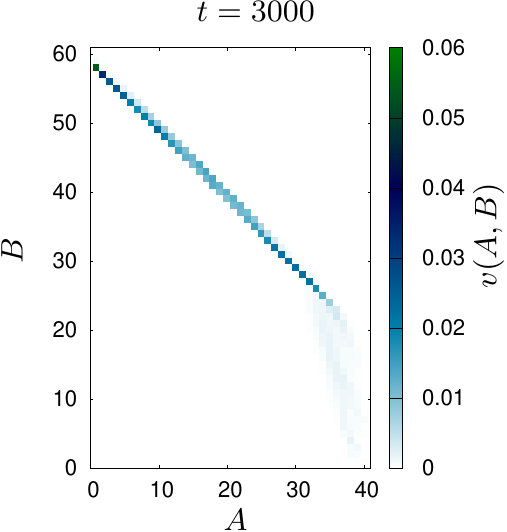}
\setfloatlink{http://www.zis.agh.edu.pl/files/random_Lx40_Ly60_alpha005.gif}
\caption{\label{vid:randomLx40Ly60alpha005}The spatial--temporal evolution of the density of actors status $v(A,B)$ for random initial state and $\alpha=0.05$, $\beta=1-\alpha$, $L_A=40$, $L_B=60$, $\delta t=10^{-5}$}
\end{video*}
%% ----------------------------------------------------------------------------
\begin{video*}
\psfrag{A}{$A$}
\psfrag{B}{$B$}
\psfrag{v(A,B)}{$v(A,B)$}
\psfrag{t=0}{$t=0$}
\psfrag{t=500}{$t=500$}
\psfrag{t=1000}{$t=1000$}
\psfrag{t=3000}{$t=3000$}
\includegraphics[height=0.23\textheight]{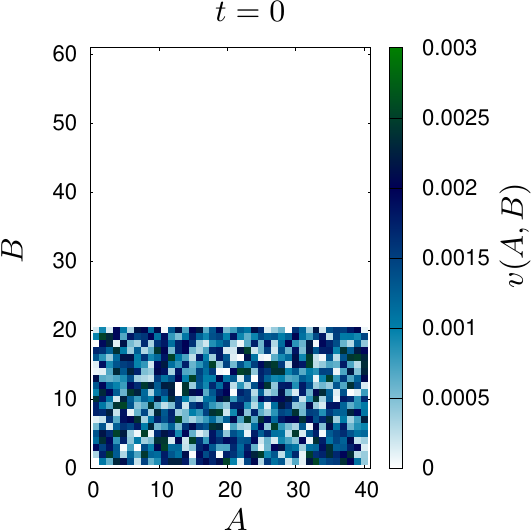}\hfill\includegraphics[height=0.23\textheight]{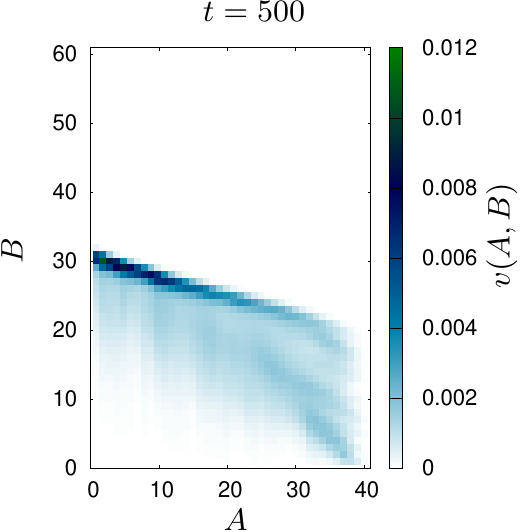}\hfill\includegraphics[height=0.23\textheight]{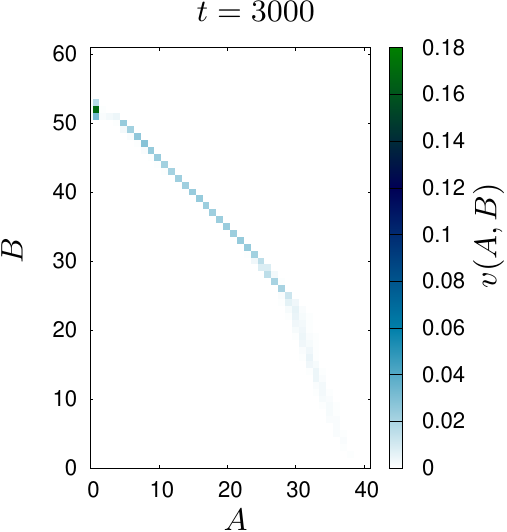}
\setfloatlink{http://www.zis.agh.edu.pl/files/random_Lx40_Ly60_alpha015.gif}
\caption{\label{vid:randomLx40Ly60alpha015}The spatial--temporal evolution of the density of actors status $v(A,B)$ for random initial state and $\alpha=0.15$, $\beta=1-\alpha$, $L_A=40$, $L_B=60$, $\delta t=10^{-5}$}
\end{video*}
%% ----------------------------------------------------------------------------
%% ----------------------------------------------------------------------------
\begin{video*}
\psfrag{A}{$A$}
\psfrag{B}{$B$}
\psfrag{v(A,B)}{$v(A,B)$}
\psfrag{t=0}{$t=0$}
\psfrag{t=500}{$t=500$}
\psfrag{t=1000}{$t=1000$}
\psfrag{t=3000}{$t=3000$}
\includegraphics[height=0.23\textheight]{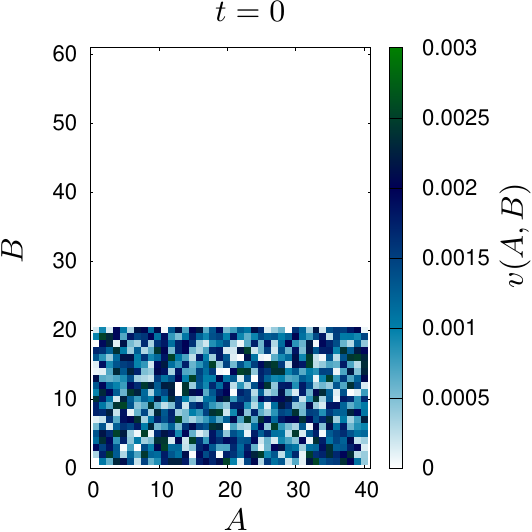}\hfill\includegraphics[height=0.23\textheight]{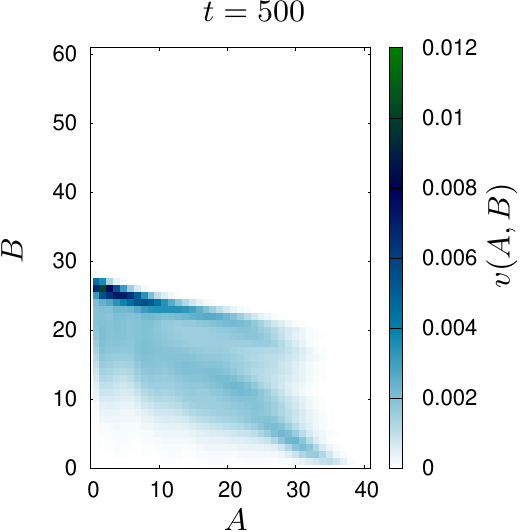}\hfill\includegraphics[height=0.23\textheight]{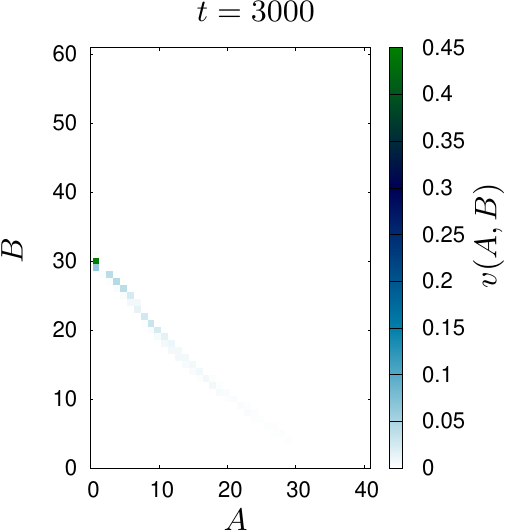}
\setfloatlink{http://www.zis.agh.edu.pl/files/random_Lx40_Ly60_alpha050.gif}
\caption{\label{vid:randomLx40Ly60alpha050}The spatial--temporal evolution of the density of actors status $v(A,B)$ for random initial state and $\alpha=0.5$, $\beta=1-\alpha$, $L_A=40$, $L_B=60$, $\delta t=10^{-5}$}
\end{video*}
%% ----------------------------------------------------------------------------
\begin{video*}
\psfrag{A}{$A$}
\psfrag{B}{$B$}
\psfrag{v(A,B)}{$v(A,B)$}
\psfrag{t=0}{$t=0$}
\psfrag{t=500}{$t=500$}
\psfrag{t=1000}{$t=1000$}
\psfrag{t=3000}{$t=3000$}
\includegraphics[height=0.23\textheight]{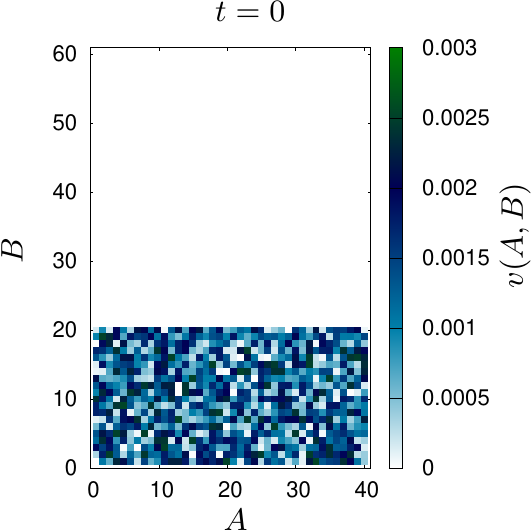}\hfill\includegraphics[height=0.23\textheight]{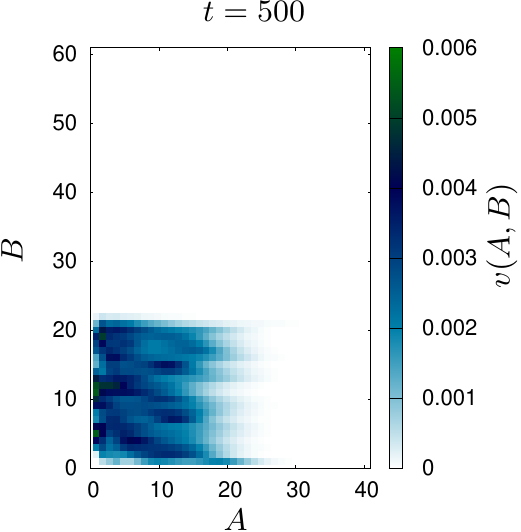}\hfill\includegraphics[height=0.23\textheight]{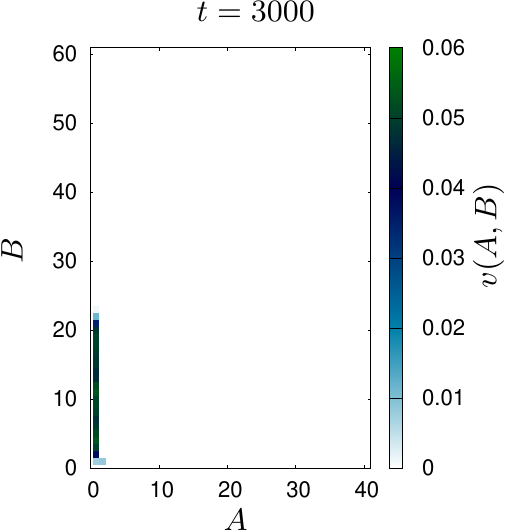}
\setfloatlink{http://www.zis.agh.edu.pl/files/random_Lx40_Ly60_alpha095.gif}
\caption{\label{vid:randomLx40Ly60alpha095}The spatial--temporal evolution of the density of actors status $v(A,B)$ for random initial state and $\alpha=0.95$, $\beta=1-\alpha$, $L_A=40$, $L_B=60$, $\delta t=10^{-5}$}
\end{video*}
%% ----------------------------------------------------------------------------

%% ############################################################################
\section{Discussion}
%% ############################################################################

The enhancement of status $B$ of actors at $(A,B)$ cell by SDS is a common action of all actors in cells $(A',B)$ such that $A'>A$. That is why the shift
along $B$-axis is the largest for actors with minimal $A$; each other actor with the same $B$ is interested in evading their rejection. This computational effect reflects tendencies to raise self-esteem of underpriviledged groups in purely symbolic way, to compensate their low social status (here: status $A$). The list of examples of SDS includes, among others:
\begin{itemize}
	\item glorification of working class in communist countries, attributing the role of dictators to proletariat \citep{manikom};
	\item strength of populism in numerous countries (see \citep{bbc} for a few examples, but the list is longer);
	\item long lists of honorary offices in the Republic of Poland in 18th century \citep{pwn};
	\item movies of action with `an ordinary cop' as main hero, brave, incorruptible and victorious (`Bullitt', `Die Hard', `Pulp Fiction' and many others);
	\item demonstrations of powerful politicians that they accept popular culture, as reproducing folk dances in public (Boris Jeltsyn \citep{bj}, Theresa May \citep{tm}).
\end{itemize}

In the first example, SDS is applied in the political international scale, as the target of the communist propaganda was the working class in the whole world. The role of the claim was to induce them to support revolution.

The second example is more contemporary and SDS in this case can be seen as less artificial, as the government there is subordinated to the conceptions of the majority about politics, no matter how misconceived they are.

The function of the phenomenon mentioned in the third example was to support the illusion of republican democracy of the Polish nobility \citep{anowak}, with its reference to ancient Rome \citep{lqc}. This example refers to one social class, one country and one epoch, but it finds counterparts in long lists of medals in the whole world: symbols of acknowledged contributions to agency of state \citep{medals}.

The fourth example refers to `cop movies', many of them seem
to be made just to attract young men and women to the profession known as low-paid, dangerous and frustrating. More about social roles of hero can be found in books by \citet{cop,cop2}.

The last example brings the myth of cultural unity of a ruler with ordinary people; eruption of such behaviours can be observed during any electoral campaign.

We claim that the model and the results presented above highlight SDS as important social phenomenon. The list of examples at the macro-level,
given above, can be added to the description of a social integration at the level of individual behavior \citep{blau1}. Taking the results
directly, the alternative of SDS is to remove the hierarchy of real status (competence and agency \citep{clr2000}) at all, which is certainly not desired by elite. On the other hand, the idea of a two-dimensional scale of status makes some questions more obvious. We can ask, for example, if academic degrees conform to $A$ or $B$ axis \citep{credentialism,cap}? A discussion of this issue exceeds the frames of this text.

To conclude, SDS is functional in society. As it was formulated by a historian, J. Elias \citet[p. 73]{gbk}: `{\em The first need of any social system
is to create incentives to make people do more work than that required by their immediate wants}'. As an incentive, the enhancement of the artificial status plays the prescribed role. More generally, the distinction between real and artificial status should be useful for further discussion.

%% ####################################################################
%% \begin{acknowledgments}
%% \end{acknowledgments}
%% ###################################################################

\bibliography{Blau}{}

\appendix

\definecolor{mygray}{rgb}{0.92,0.97,0.92}
\lstset{backgroundcolor=\color{mygray},commentstyle=\itshape\color{blue},stringstyle=\color{teal}}

\begin{widetext}
\begin{lstlisting}[language={[95]Fortran},frame=single,numbers=left,numberstyle=\tiny,basicstyle=\footnotesize,stepnumber=1,breaklines=true,caption={Fortran95 code allowing for reproduction of Videos~\ref{vid:hierar_Lx20_Ly20_alpha005}--\ref{vid:randomLx40Ly60alpha095}},label=lst:main]
!! Paradox of integration---Dynamics of two-dimensional status
!! K. Malarz <malarz@agh.edu.pl>
!! Sun, 3 Feb 2019, 16:51:24 CET

!! ================================================================
module settings
!! ================================================================
implicit none

integer, parameter :: Lx=40, Ly=60, N=3e+8
real*8, parameter :: dt=0.00001d0, alpha=0.95d0, beta=1.0d0-alpha
end module settings

!! ################################################################
program Blau
!! ################################################################
use settings
implicit none
integer ii, xx, x, y
real*8, dimension (0:Lx+1,0:Ly+1) :: v, vnew
real*8, dimension (4) :: w
real*8 :: norm

print '(A3,2A10,3A10)','###','Lx','Ly','alpha','beta','dt'
print '(A3,2I10,3F10.6)','###',Lx,Ly,alpha,beta,dt

v=0.0d0

!! ================================================================
! hierarchical initial state
!! ================================================================
!do x=1,Lx
!  v(x,1)=1.0d0/(1.0d0*Lx)
!enddo
!! ================================================================

!! ================================================================
! homogenous initial state
!! ================================================================
! do x=1,Lx
! do y=1,Lx/2
!   v(x,y)=1.0d0/(0.5d0*Lx*Lx)
! enddo
! enddo
!! ================================================================

!! ================================================================
! random initial state
!! ================================================================
do x=1,Lx
do y=1,Lx/2
  v(x,y)=1.0d0*rand()
enddo
enddo
norm=sum(v)
v=v/norm
!! ================================================================


norm=sum(v)
ii=0
print *,'# ',ii*dt,norm

     do x=1,Lx
     do y=1,Ly
       print *,x,y,v(x,y)
     enddo
     enddo

do ii=1,N

  do x=1,Lx
  do y=1,Ly

    w(1)=0.0d0
    do xx=1,x-1
      w(1)=w(1)+v(xx,y)
    enddo
    w(1)=alpha*w(1)

    w(2)=0.0d0
    do xx=x+1,Lx
      w(2)=w(2)+v(xx,y)
    enddo
    w(2)=beta*w(2)

    w(3)=0.0d0
    do xx=1,x
      w(3)=w(3)+v(xx,y)
    enddo
    w(3)=alpha*w(3)

    w(4)=0.0d0
    do xx=x+1,Lx
      w(4)=w(4)+v(xx,y-1)
    enddo
    w(4)=beta*w(4)

    vnew(x,y)=v(x,y)+dt*(-v(x,y)*(w(1)+w(2))+v(x+1,y)*w(3)+v(x,y-1)*w(4))

  enddo
  enddo

  do x=1,Lx
  do y=1,Ly
    v(x,y)=vnew(x,y)
  enddo
  enddo

   if(mod(ii,500000).eq.0) then
     norm=sum(v)
     print *,''
     print *,'# ',ii*dt,norm
     do x=1,Lx
     do y=1,Ly
       print *,x,y,v(x,y)
     enddo
     enddo
   endif
enddo

end program Blau
\end{lstlisting}
\end{widetext}

\end{document}